\documentclass
[aps,prx,superscriptaddress,reprint,showpacs,amssymb,amsmath,floatfix,citeautoscript,longbibliography]{revtex4-2}

\usepackage{tabulary} 
\usepackage{tabularx} 
\usepackage{graphicx} 
\usepackage{dcolumn} 
\usepackage{bm} 

\usepackage{ulem}
\usepackage[]{color}

\begin{document}

\title{Dual nature of 5$f$ electrons in the isostructural U$M_2$Si$_2$ family: from antiferro- to Pauli paramagnetism via  hidden order}

\author{Andrea~Amorese}
 \affiliation{Institute of Physics II, University of Cologne, Z{\"u}lpicher Stra{\ss}e 77, 50937 Cologne, Germany}
 \affiliation{Max Planck Institute for Chemical Physics of Solids, N{\"o}thnitzer Stra{\ss}e 40, 01187 Dresden, Germany}
\author{Martin~Sundermann}
 \affiliation{Institute of Physics II, University of Cologne, Z{\"u}lpicher Stra{\ss}e 77, 50937 Cologne, Germany}
 \affiliation{Max Planck Institute for Chemical Physics of Solids, N{\"o}thnitzer Stra{\ss}e 40, 01187 Dresden, Germany}
\author{Brett~Leedahl}
	\affiliation{Max Planck Institute for Chemical Physics of Solids, N{\"o}thnitzer Stra{\ss}e 40, 01187 Dresden, Germany}
\author{Andrea~Marino}
 \affiliation{Max Planck Institute for Chemical Physics of Solids, N{\"o}thnitzer Stra{\ss}e 40, 01187 Dresden, Germany}
 \affiliation{Dipartimento di Fisica, Politecnico di Milano, Piazza Leonardo da Vinci 32, I-20133 Milano, Italy}
\author{Daisuke~Takegami}
 \affiliation{Max Planck Institute for Chemical Physics of Solids, N{\"o}thnitzer Stra{\ss}e 40, 01187 Dresden, Germany}
\author{Hlynur~Gretarsson}
 \affiliation{Max Planck Institute for Chemical Physics of Solids, N{\"o}thnitzer Stra{\ss}e 40, 01187 Dresden, Germany}
 \affiliation{PETRA III, Deutsches Elektronen-Synchrotron (DESY), Notkestra{\ss}e 85, 22607 Hamburg, Germany}
\author{Andrei~Hloskovsky}
 \affiliation{PETRA III, Deutsches Elektronen-Synchrotron (DESY), Notkestra{\ss}e 85, 22607 Hamburg, Germany}
\author{Christoph~Schl\"uter}
 \affiliation{PETRA III, Deutsches Elektronen-Synchrotron (DESY), Notkestra{\ss}e 85, 22607 Hamburg, Germany}
\author{Maurits~W.~Haverkort}
  \affiliation{Institute for Theoretical Physics, Heidelberg University, Philosophenweg 19, 69120 Heidelberg, Germany}
\author{Yingkai~Huang}
  \affiliation{van der Waals-Zeeman Institute, University of Amsterdam, 1098 XH Amsterdam, The Netherlands}
\author{Maria~Szlawska}
  \affiliation{Institute of Low Temperature \& Structure Research, Polish Academy of Science, Wroclaw, Poland}
\author{Dariusz~Kaczorowski}
  \affiliation{Institute of Low Temperature \& Structure Research, Polish Academy of Science, Wroclaw, Poland}
\author{Sheng Ran}
  \altaffiliation{Present address: Physics department, Washington University in St. Louis, St. Louis, Missouri, USA 63130.}
  \affiliation{Department of Physics, University of California, San Diego, La Jolla, California, USA}
\author{M. Brian Maple}
  \affiliation{Department of Physics, University of California, San Diego, La Jolla, California, USA}
\author{Eric D. Bauer}
  \affiliation{Los Alamos National Laboratory, Los Alamos, New Mexico 87545, USA}
\author{Andreas~Leithe-Jasper}
	\affiliation{Max Planck Institute for Chemical Physics of Solids, N{\"o}thnitzer Stra{\ss}e 40, 01187 Dresden, Germany}
\author{Philipp Hansmann}
	\altaffiliation{Present address: Department of Physics, University of Erlangen - Nuremberg, 91058 Erlangen, Germany}
	\affiliation{Max Planck Institute for Chemical Physics of Solids, N{\"o}thnitzer Stra{\ss}e 40, 01187 Dresden, Germany}
\author{Peter~Thalmeier}
	\affiliation{Max Planck Institute for Chemical Physics of Solids, N{\"o}thnitzer Stra{\ss}e 40, 01187 Dresden, Germany}
\author{Liu~Hao~Tjeng}
	\affiliation{Max Planck Institute for Chemical Physics of Solids, N{\"o}thnitzer Stra{\ss}e 40, 01187 Dresden, Germany}
\author{Andrea~Severing}
  \affiliation{Institute of Physics II, University of Cologne, Z{\"u}lpicher Stra{\ss}e 77, 50937 Cologne, Germany}
  \affiliation{Max Planck Institute for Chemical Physics of Solids, N{\"o}thnitzer Stra{\ss}e 40, 01187 Dresden, Germany}	
\date{\today}

\begin{abstract}
Using inelastic x-ray scattering beyond the dipole limit and hard x-ray photoelectron spectroscopy we establish the dual nature of the U $5f$ electrons in U$M_2$Si$_2$ ($M$ = Pd, Ni, Ru, Fe), regardless of their degree of delocalization. We have observed that the compounds have in common a local atomic-like state that is well described by the U $5f^2$ configuration with the $\Gamma_1^{(1)}$ and $\Gamma_2$ quasi-doublet symmetry. The amount of the U 5$f^3$ configuration, however, varies considerably across the U$M_2$Si$_2$ series, indicating an increase of U\,5$f$ itineracy in going from $M$\,=\,Pd to Ni to Ru, and to the Fe compound. The identified electronic states explain the formation of the very large ordered magnetic moments in UPd$_2$Si$_2$ and UNi$_2$Si$_2$, the availability of orbital degrees of freedom needed for the hidden order in URu$_2$Si$_2$ to occur, as well as the appearance of Pauli paramagnetism in UFe$_2$Si$_2$. A unified and systematic picture of the U$M_2$Si$_2$ compounds may now be drawn, thereby providing suggestions for new experiments to induce hidden order and/or superconductivity in U compounds with the tetragonal body-centered ThCr$_2$Si$_2$ structure.
\end{abstract}

\maketitle

In heavy fermion compounds the intricate interplay between the $f$ and conduction electrons has a large impact on ground state properties\,\cite{Floquet2005,Thalmeier2005,Coleman2007,Hilbert2007,Khomskii2010,Stockert2012,White2015}. Herein we study uranium $5f$ systems with the U$M_2$Si$_2$ composition that crystallize in the tetragonal body-centered ThCr$_2$Si$_2$ structure whereby $M$ denotes a transition metal. Members of this family exhibit a strong $a$-$c$-axis magnetic anisotropy and several of them show long-range magnetic order (e.g., $M$ = Pd, Ni) or remain Pauli paramagnetic (e.g. $M$ = Fe) down to low temperatures\,\cite{Ptasiewicz1981,Buschow1986,Palstra1986,Lin1991,Endstra1993b,Shemirani1993,Svoboda2004,Szytuka2007,Plackowski2011}. URu$_2$Si$_2$ is special; it undergoes two transitions, one into an ordered state at 17.5\,K with a considerable loss of entropy ($\approx$\,0.2\,$R$ln2) and a second one at 1.5\,K into a superconducting phase\,\cite{Palstra1985,Schlabitz1986,Maple1986}. Below 17.5\,K, ordered magnetic moments of 0.03\,$\mu_B$ have been measured\,\cite{Broholm1987,Niklowitz2010}, but the moment is too small to account for the loss of entropy. Therefore, it is believed that the phase below 17.5\,K is an electronically ordered state but with an order parameter that is yet unknown and continues to be heavily debated to this day, see Refs.\,\cite{Oppeneer2010,Mydosh2011,Ikeda2012,Mydosh2014,Kung2015,Mydosh2020} and references therein. This is the famous \textsl{hidden order} phase (HO). The application of pressure, however, suppresses the HO phase and a large moment antiferromagnetic (LMAFM) phase develops. At about 5\,kbar the ordered magnetic moment rises discontinuously from 0.03 to about 0.4\,$\mu_B$\,\cite{Amitsuka2007,Niklowitz2010}. Also, magnetic field acts to suppress the HO state and instead a spin density wave has been observed\,\cite{Knafo2015}.

In these uranium systems, the $5f$ electrons are crucial for the ground state formation. This situation raises the question whether a systematic picture can be developed that takes into account both correlation effects and band formation with the $5f$ states, and at the same time explains consistently the widely varying properties of the U$M_2$Si$_2$ compounds. One of the most pressing issues is whether local or atomic-like states can survive the band formation in such metallic systems, or in other words, whether it is meaningful at all to develop models that have atomic multiplet states as a starting point.  
Otherwise one may be better off using band theory based methods\,(see Refs.\,\cite{Oppeneer2010,Mydosh2011,Mydosh2014,Mydosh2020} and references therein). Very recently non-resonant inelastic x-ray scattering (NIXS, or x-ray Raman scattering) beyond the dipole limit revealed that local atomic multiplet states can be identified in URu$_2$Si$_2$ \cite{Sundermann2016}, which is quite surprising since itineracy and Fermi surface effects do play a role in the HO transition\,\cite{Santander2009,Meng2013}. It is now important to investigate whether the other members of the U$M_2$Si$_2$ family show multiplets, and if so, whether the atomic multiplet states are the same or different across the isostructural family.

For our present study, we selected $M$ = Pd, Ni, Ru, Fe, with Fe\,(Pd) being isoelectronic with Ru\,(Ni).  The UPd$_2$Si$_2$ and UNi$_2$Si$_2$ order antiferromagnetically (AF) at $T_N$\,=\,136 and 124\,K, respectively\,\cite{Svoboda2004,Plackowski2011} with very large ordered moments; 3.37\,\cite{Ptasiewicz1981} and 2.3\,$\mu_B$\,\cite{Shemirani1993} have been reported for UPd$_2$Si$_2$ and 2.7\,$\mu_B$, for UNi$_2$Si$_2$\,\cite{Lin1991}. The ordered magnetic moments are, like in URu$_2$Si$_2$ under pressure, aligned along the $c$ direction with propagation vectors Q\,=\,(0,0,2$\pi$/c) at the ordering transition. URu$_2$Si$_2$ is the HO compound exhibiting superconductivity, and UFe$_2$Si$_2$ is a Pauli paramagnet (PP) down to the lowest temperatures\,\cite{Palstra1986,Endstra1993b,Szytuka2007}. We thus cover a wide range of physical properties while keeping the same U-Si framework and crystal structure ($I4/mmm$). We will apply NIXS at the U $O_{4,5}$ edges ($5d\rightarrow5f$) in order to determine the presence and symmetry of possible localized $5f$ states. For measuring the degree of delocalization, we will utilize U\,4$f$ core-level photoelectron spectroscopy (PES). This is one of the most powerful spectroscopic methods to study hybridization effects in U compounds\,\cite{Grassmann1990,Fujimori2012}. Here we apply the hard x-ray version of PES (HAXPES) in order to make use of the larger probing depth and thus to ensure that the signal is representative for the bulk material. 

Our objective is to establish whether the so-called \textsl{dual nature of f-electrons} model proposed for the description of both the antiferromagnetic order and heavy fermion properties of UPt$_3$\,\cite{Zwicknagl2002,Zwicknagl2003a} and UPd$_2$Al$_3$\,\cite{Thalmeier2002,Zwicknagl2003} is also a feasible concept to capture the low-energy electronic structure of the hidden order and Pauli paramagnetic members of the U$M_2$Si$_2$ family, and not only of antiferromagnetic members such as UPt$_2$Si$_2$\,\cite{Lee2018}. If so, we may be able to draw a systematic picture which in turn can be used to provide a solid basis for the realistic modeling of the HO transition, and to point out further experiments to induce HO or superconductivity in other members of the U$M_2$Si$_2$ family.

\section{Results}

\subsection{Ground state symmetry with NIXS}
In NIXS the directional dependence of the double differential cross-section gives insight to the orbital anisotropy of the ground state, similar to the linear dichroism in x-ray absorption (XAS). Here the direction of the momentum transfer $\vec{q}$, acts similarly to the direction of the electric field vector in XAS. The size of the momentum transfer $\left|\vec{q}\right|$ makes the important difference; for large momentum transfers NIXS is governed by multipole selection rules while XAS is governed by dipole selection rules. The multipole scattering of the U O$_{4,5}$ edge is more excitonic so that a local atomic approach can be used for a quantitative analysis of the spectra\,\cite{Caciuffo2010}. This and further explanations why the NIXS U O$_{4,5}$ is sensitive to the symmetry of the U configuration that is lowest in energy can be found in Ref.\,\cite{Sundermann2016} and references therein. Further credibility of the U O$_{4,5}$ NIXS method is given in Ref.\,\cite{Sundermann2018a} which shows that NIXS confirms the ground state symmetry of UO$_2$ that was determined with inelastic neutron scattering.

\begin{figure}
    \centerline{\includegraphics[width=0.9\columnwidth]{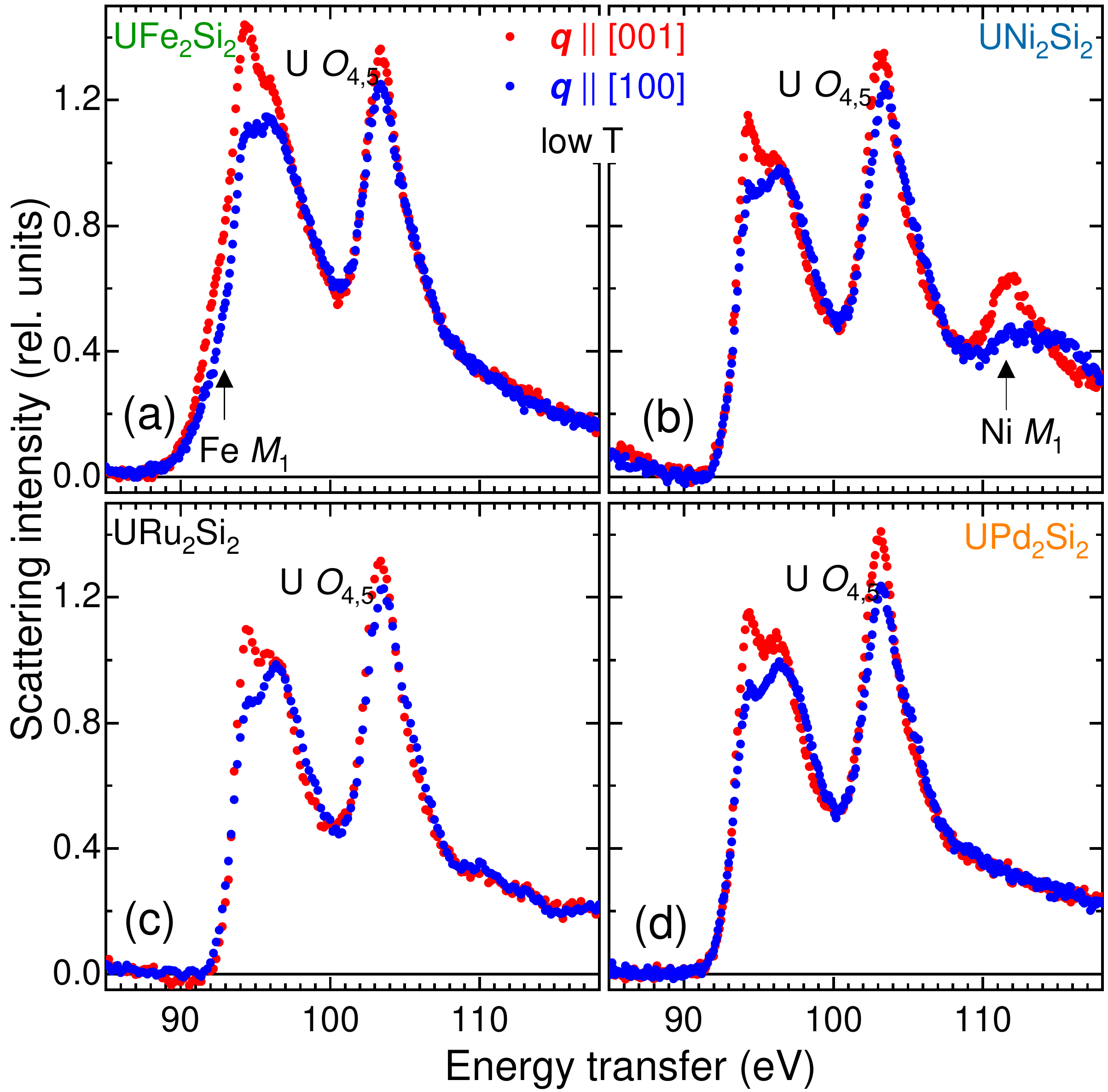}}
    \caption{(Normalized and background corrected experimental NIXS data I$_{\vec{q}\|[100]}$ (blue dots) and I$_{\vec{q}\|[001]}$ (red dots) at the U $O_{4,5}$ edges (5$d\rightarrow$5$f$) at $T$\,$<$ 15\,K. The URu$_2$Si$_2$ data in panel (c) are adapted from Ref.\,\cite{Sundermann2016}. The size of the data points represents the statistical error.}
    \label{NIXS_O45}
\end{figure}

The dominant NIXS signal arises from Compton scattering and the core-level excitations appear as spikes on top (see Fig.\,\ref{Compton} in Appendix). Not all core-levels have a sizable cross-section at $\left|\vec{q}\right|$\,=\,9.6\,\AA$^{-1}$, but the U $O_{4,5}$ core level at 100\,eV energy transfer is distinctly visible in all the spectra. The broad Compton background was used for normalizing the spectra of different $\vec{q}$ directions of one compound. In the second step the data of the U $O_{4,5}$ edges of the different compounds were normalized to each other using the isotropic spectra that are constructed from directional dependent $O_{4,5}$ edge data in Fig.\,\ref{NIXS_O45} as $I_{\text{iso}}$\,=\,(2$\cdot I_{\vec{q}\|[100]}$\,+\,$I_{\vec{q}\|[001]}$)/3 \footnote{The isotropic spectrum (2$\cdot{I}_{\vec{q}\|[100]}$\,+\,${I}_{\vec{q}\|[001]}$)/3 is a pseudoisotropic spectrum in the beyond dipole limit but the deviations are minor, see PhD thesis Sundermann\,\cite{Sundermann_PhD}}. 

Figure\,\ref{NIXS_O45} shows the $O_{4,5}$ edge data for $T$\,$<$15\,K of UFe$_2$Si$_2$ (a), UNi$_2$Si$_2$ (b), URu$_2$Si$_ 2$(c) and UPd$_2$Si$_2$ (d) for $\vec{q}\|$[100] (blue) and $\vec{q}\|$[001] (red) measured with energy steps of 0.1\,eV (0.2\,eV for $M$\,=\,Ru). The size of the data points reflects the statistical error bars. The data were  normalized (see above) and a linear background was subtracted. Lastly, the URu$_2$Si$_2$ data were reproduced from Ref.\,\cite{Sundermann2016}. 

All four spectra in Fig.\,\ref{NIXS_O45} exhibit a clear directional dependence (dichroism) and the similarities in magnitude and line shape of the spectra are apparent. The differences between the four compounds are only due to the appearance of the dipole forbidden $M_1$ edges (3$s\rightarrow$3$d$) of the Ni sample at 112\,eV and of the Fe sample at 91\,eV. For Ni the $M_1$ edge lies above the higher energy branch of the U $O_{4,5}$ edge but for Fe it coincides with the lower energy branch of the U $O_{4,5}$ edge. These $M_1$ edges also exhibit a dichroism\,\cite{Yavas2019} so that in case of UFe$_2$Si$_2$ we mainly rely on the directional dependence of the higher energy branch of the U\,$O_{4,5}$ edge at 103\,eV. Otherwise, the shape of the U\,$O_{4,5}$ edges seem fairly robust and independent of the compound under investigations.

In Ref.\,\cite{Sundermann2016} we showed that the multiplet structure of the isotropic NIXS spectrum of URu$_2$Si$_2$ is well reproduced with a U$^{4+}$ $5f^2$ \textsl{ansatz} in its $J$\,=\,4 ground-state multiplet. Higher multiplets do not contribute to the ground state consistent with the fact that the expected crystal-field splittings as well as the Kondo scale of URu$_2$Si$_2$ are much smaller than the spin-orbit splitting of the order of 550-600\,meV\,\cite{Butorin2000,Wray2015}. The similarity of the NIXS spectra thus indicates that all the other compounds with $M$\,=\,Fe, Ni, and Pd show the presence of atomic-like multiplet states, and that these are also 5$f^2$ based. The presence of hybridization and covalency effects is taken into account effectively by the reduction factors of 50\% of the Slater integrals 5$f$-5$f$ and 5$d$-5$f$ (see section Materials and Methods C.). 

\begin{figure*}[t]
    \includegraphics[width=0.45\textwidth]{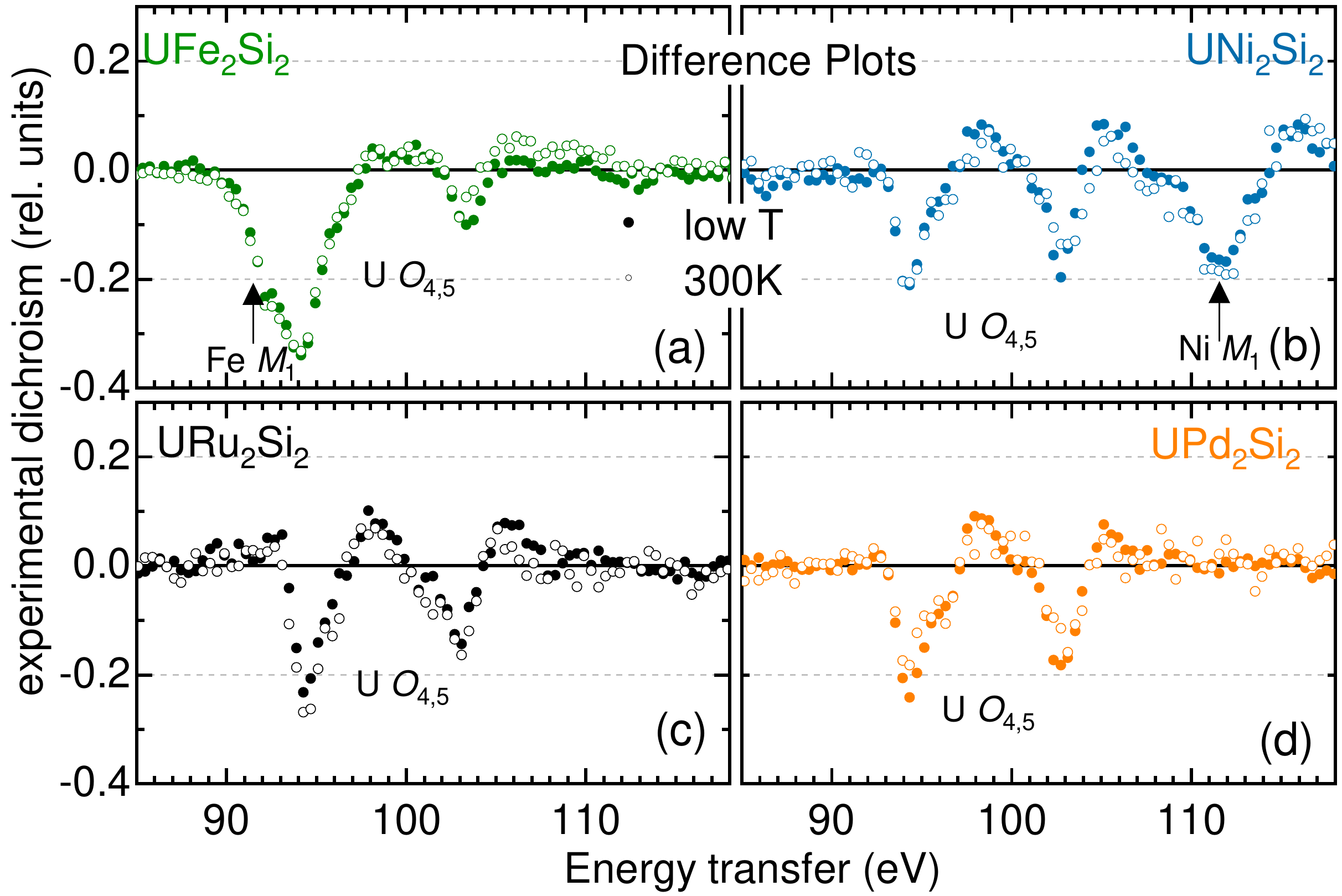}
    \includegraphics[width=0.45\textwidth]{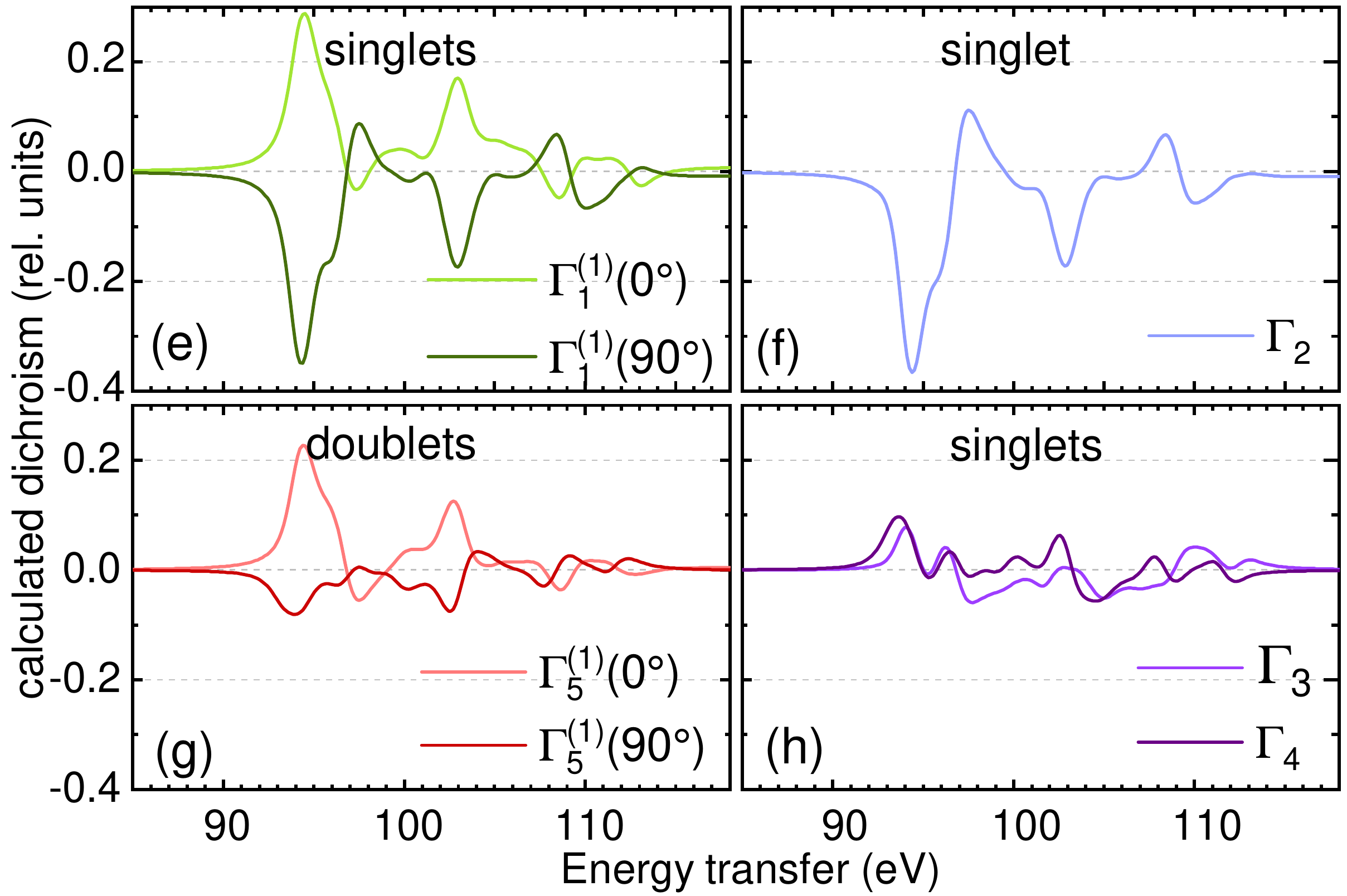}
		\caption{Difference plots I$_{\vec{q}\|[100]}$-I$_{\vec{q}\|[001]}$ (dichroism) (green for $M$\,=\,Fe, blue for Ni, black for Ru, orange for Pd) of U$M_2$Si$_2$. Full circles represent the dichroism at $T$\,$<$ 15\,K and open circles at $T$\,=\,300\,K. The size of the data points represents the statistical error. (e)-(f) calculated dichroism for the seven crystal-field states; note the dichroism of $\Gamma_{1,5}^{(1)}$(0$^{\circ}$) is equal to the dichroism of $\Gamma_{1,5}^{(2)}$(90$^{\circ}$) and the one of $\Gamma_{1,5}^{(1)}$(90$^{\circ}$) is equal to the one of $\Gamma_{1,5}^{(2)}$(0$^{\circ}$). In panels (a) and (c) the extremes are shown for the $\Gamma_1^{1,2}$($\theta$) singlet and $\Gamma_5^{(1,2)}$($\phi$) doublet states with mixed $J_z$ (see eq.(1), (2), (6) and (7)).}
    \label{dichro}
\end{figure*}

Fig.\,\ref{dichro}\,(a)-(d) shows the difference spectra (dichroism) of the directional dependent NIXS data. Given the similarity of the NIXS spectra in Fig.\,\ref{NIXS_O45}\,(a)-(d) it does not come as a surprise that the dichroisms of antiferromagnetic UNi$_2$Si$_2$ and UPd$_2$Si$_2$ agree as well with the one of URu$_2$Si$_2$ in the region of the U $O_{4,5}$ edge. For the Pauli paramagnet UFe$_2$Si$_2$, we also find that the dichroism agrees well with the one the other compounds although we have to restrict the comparison to the U signal at 103\,eV in order to avoid the contribution of Fe $M_1$.    

The directional dependence in the spectra is due to the crystal-field splitting of the Hund's rule ground state of the U 5$f^2$ configuration. It has a total angular momentum $J$\,=\,4 and is split by the tetragonal ($D_{4h}$-symmetry) crystalline electric field (CEF) into five singlets and two doublets. $J$ remains a good quantum number even in the intermediate coupling regime of uranium so that the CEF wave functions can be written in terms of $J_z$ (see eqs. (1)-(7)). 

\begin{eqnarray}
\Gamma_1^{(1)}(\theta) = \cos(\theta) \, | 0 \rangle + \sin(\theta) \sqrt{\frac{1}{2}} ( | +4 \rangle + | -4 \rangle )  \\
\Gamma_1^{(2)}(\theta) = \sin(\theta) \, | 0 \rangle - \cos(\theta) \sqrt{\frac{1}{2}} ( | +4 \rangle + | -4 \rangle )  \\
\Gamma_2 = \sqrt{\frac{1}{2}} ( | +4 \rangle - | -4 \rangle ) \\
\Gamma_3 = \sqrt{\frac{1}{2}} ( | +2 \rangle + | -2 \rangle ) \\
\Gamma_4 = \sqrt{\frac{1}{2}} ( | +2 \rangle - | -2 \rangle ) \\
\Gamma_5^{(1)}(\phi) = \cos(\phi) \, | \mp 1 \rangle + \sin(\phi) \, | \pm 3 \rangle \\
\Gamma_5^{(2)}(\phi) = \sin(\phi) \, | \mp 1 \rangle - \cos(\phi) \, | \pm 3 \rangle  
\end{eqnarray}

We now calculate the dichroism of the seven CEF states (see Fig.\,\ref{dichro}\,(e)-(h)) with the full multiplet code \textsl{Quanty}\,\cite{Haverkort2016} using the same parameters as in Ref.\,\cite{Sundermann2016} (see Appendix). For the mixed singlet states $\Gamma_1^{(1,2)}$($\theta$) (eq.\,1 and 2) in panel (e) and the doublet states (eq.\,6 and 7) in panel (g) the extreme dichroisms are given for $\theta$ and $\phi$ equal to 0 and 90$^{\circ}$. For other values of $\theta$ and $\phi$, i.e., in case of $J_z$ mixtures the dichroism falls in between the dark and light green (red) lines. In contrast, the dichroism of the $\Gamma_2$, $\Gamma_3$ and $\Gamma_4$ singlet states in panel (f) and (h), respectively, are given by single lines. 

The comparison of the experimental directional dependencies in Fig.\,\ref{dichro}\,(a)-(d) and the dichroism of the seven crystal-field states in Fig.\,\ref{dichro}\,(e)-(h) shows immediately that for all four U$M_2$Si$_2$ compounds investigated, only the $\Gamma_1^{(1)}$($\approx$\,90$^{\circ}$) and the $\Gamma_2$ have the correct sign and magnitude to reproduce the difference spectra\,\footnote{About 10\% (75$^{\circ}\leq\theta\leq$105$^{\circ}$) of $| 0 \rangle$ entering the $\Gamma_1^{(1)}$ is possible, giving a slight reduction of the magnitude of the dichroism.}. In fact, the experimentally observed magnitude is so large that it excludes any other state, also the $\Gamma_5^{(1,2)}$ that was deduced from O-edge XAS measurements on URu$_2$Si$_2$ by Wray\textit{ et al}.\,\cite{Wray2015}; for any $J_z$ admixture its directional dependence is either too weak or has the wrong sign (see Fig.\,2\,(g)). Note, in spectroscopy, it is always possible to lose dichroism due to, e.g., surface issues in a surface sensitive technique. NIXS, however, is bulk sensitive and we detect almost the largest possible directional dependence. Finding the same large dichroism in all four compounds, we can safely conclude that it is the singlet $\Gamma_1^{(1)}$($\approx$90$^{\circ}$) or the singlet $\Gamma_2$ or, as will be explained below, a \textsl{quasi-doublet} made up of the two that determines the local symmetry in the ground state, and that the four compounds have this result in common. 

Additional data were taken at 300\,K in order to search for the population of CEF excited states. The open circles in Fig.\,\ref{dichro}\,(a)-(d) represent the directional dependence of the 300\,K data. We find that the temperature effect is negligible within the error bars of the experiment for all compounds when comparing the open ($T$ \,=\,300\,K) with the full circles ($T$\,$<$\,15\,K). This means that the excited states do not get thermally populated and are quite far away from the $\Gamma_1^{(1)}$($\approx$\,90$^{\circ}$) or the $\Gamma_2$ singlet, or their quasi-doublet ground state. In Ref.\,\cite{Sundermann2016}, we had estimated from the lack of temperature dependence that the states with weak dichroism like the $\Gamma^{(1)}_5$(90$^{\circ}$), the $\Gamma_3$ and $\Gamma_4$ must be higher than 150\,K (13 meV), whereas states with stronger opposite anisotropy must be even higher in energy (compare Fig.\,\ref{dichro}\,(e)-(h)). Only in the case of UPd$_2$Si$_2$ the directional dependence seems to have decreased slightly with rising $T$, hinting towards a smaller CEF splitting with respect to the other compounds. 

\subsection{Relative 5$f$ electron count with HAXPES}

\begin{figure}[t]
    \centerline{\includegraphics[width=0.9\columnwidth]{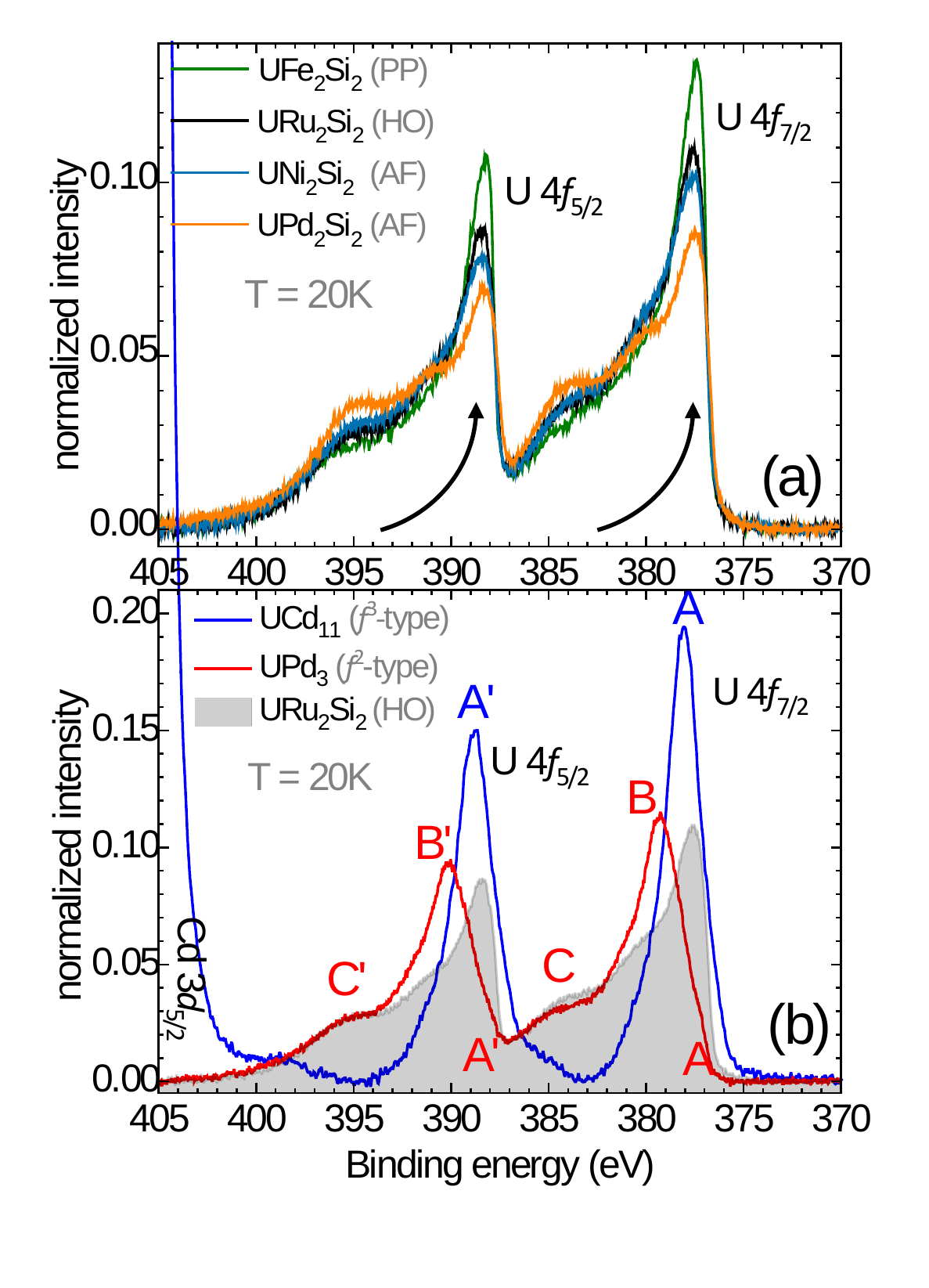}}
    \caption{Background corrected U 4$f$ core level hard x-ray photoelectron spectroscopy (HAXPES) data (h$\nu$\,=\,5945\, eV) normalized to the integrated intensity. (a) of UFe$_2$Si$_2$ (green), URu$_2$Si$_2$ (black), UNi$_2$Si$_2$ (blue) and UPd$_2$Si$_2$ (orange), and (b) of the reference compounds UCd$_{11}$ (blue line) and UPd$_3$ (red line) with URu$_2$Si$_2$ (gray filling) for comparison. Note the expanded y-scale in (b).}
    \label{HAXPES}
\end{figure}

Figure\,\ref{HAXPES}\,(a) shows the U\,4$f$ core-level HAXPES data of UFe$_2$Si$_2$, URu$_2$Si$_2$, UNi$_2$Si$_2$, and UPd$_2$Si$_2$ at $T$\,=20\,K after subtracting an integral-type (Shirley) background\,\cite{Hufner} and normalization to the integrated intensity of the U 4$f$ core-level emission lines. The U 4$f$ core level is spin-orbit split by about 10.8\,eV ($J$\,=\,5/2 and 7/2) and the intensity ratio of the spectral weights assigned to the U\,4$f_{5/2}$ and U\,4$f_{7/2}$ turns out to be about 0.8 for all four compounds which agrees well with the expected value of 6/8\,=\,0.75. The U 4$f$ core-level data consist of the superposition of several U configurations, each with its own multiplet structure. We may crudely describe the spectra with a triple peak structure at 377.5\,(388.3), 380\,(390.8), and 384\,(394.5)\,eV for U 4$f_{7/2}$\,(U 4$f_{5/2}$) (see Fig.\,\ref{HAXPES}\,(a)). 

A systematic change becomes apparent when comparing the 4$f$ core-level spectra of the four compounds; from Pd\,$\rightarrow\,$Ni\,$\rightarrow$\,Ru\,$\rightarrow$\,Fe the higher energy spectral weight at 384\,(394.8)\,eV of the U\,4$f_{7/2}$\,(U\,4$f_{5/2}$) core level loses spectral weight to the benefit of the peak at 377.5\,(388.5)\,eV. The relative change in spectral weights cannot be due to different crystal structures, different multiplets or different ground-state symmetries because the four compounds are isostructural and it was shown in the previous section that all four compounds have the same ground-state symmetry arising out of an U$^{4+}$\,5$f^2$ configuration. It can therefore only be due to a change in the 5$f$-shell occupation. To be more specific, it must be due to a successive increase of the number of $f$ electrons in the 5$f$-shell in accordance with the sequence Pd\,$\rightarrow\,$Ni\,$\rightarrow$\,Ru\,$\rightarrow$\,Fe. 

The justification for this interpretation of $increasing$ 5$f$ shell filling from $M$\,=\,Pd to Fe is given in Fig.\,\ref{HAXPES}\,(b) which shows the U\,4$f$ core-level HAXPES data of UCd$_{11}$ (blue line) and UPd$_3$ (red line). Again the data are normalized to the integrated intensity (note the larger y-scale in comparison to panel (a) to accommodate the strong UCd$_{11}$ signal). UCd$_{11}$ is an example for an intermetallic U compound that has adopted the 5$f^3$ configuration\,\cite{Tobin2019} and it shows a simple U\,4$f$ core-level spectrum with peaks at A (U\,4$f_{7/2}$) and A' (U 4$f_{5/2}$). Also the isotropic NIXS spectrum of  UCd$_{11}$ is that of a local U\,$f^3$, showing the typical shift in energy for a decrease in valence by one\,\cite{Agrestini2018} and a different lineshape than the $f^2$ configuration (see Fig.\,\ref{isotropic}in Appendix).  UPd$_3$, on the other hand is an intermetallic U compound that is quite localized and well described by the U\,5$f^2$ configuration\,\cite{McEwen2007}. It shows a pronounced double peak structure B (B') and C (C') for U 4$f_{7/2}$\,(U 4$f_{5/2}$) with some minor, third contribution A (A'), very much in agreement with Fujimori \textsl{et al.}\cite{Fujimori2012}. 

For a better comparison we overlaid the spectrum of URu$_2$Si$_2$ in Fig.\,\ref{HAXPES}\,(b). This clearly reveals that URu$_2$Si$_2$ is intermediate valent because the spectrum contains the A,\,B,\,C (A',\,B',\,C') structure of the U\,5$f^2$ and U\,5$f^3$ features. The peak positions in UPd$_3$ and UCd$_{11}$ are not precisely the same as in U$M_2$Si$_2$ compounds which can be attributed to the different chemical environment of the U atoms.  We further know from a configuration interaction analysis of PES data of e.g. cerium compounds\,\cite{Fuggle1983,Gunnarsson1983,Kotani1988} that the higher $f$-shell filling has an over-proportional higher spectral weight, so that, without attempting a quantitative analysis, we can further state that the amount of the 5$f^2$ configuration in the initial state must be significant. Another look at Fig.\,\ref{HAXPES}\,(a) lets us then conclude that the U 5$f^3$ contribution increases successively from the two antiferromagnets UPd$_2$Si$_2$ and UNi$_2$Si$_2$ to the hidden order compound URu$_2$Si$_2$, and to the Pauli paramagnet UFe$_2$Si$_2$.

\section{Discussion}
In NIXS all four compounds exhibit multiplets that are well described with the U\,5$f^2$ local symmetry i.e. the multiplets survive even the itineracy in the Pauli paramagnetic state in UFe$_2$Si$_2$. Hence, irrespective of the degree of itineracy, the U\,5$f^2$ configuration determines the local symmetry. This gives credibility to our previous findings of U 5$f^2$ multiplets in the hidden order compound URu$_2$Si$_2$\,\cite{Sundermann2016}. Together with the local symmetry contributions, all four compounds have to be classified as intermediate valent; their ground states are mixtures of the U\,5$f^2$ and U\,5$f^3$ configurations. The overall presence of multiplets implies that \textsl{the dual nature of $f$ electrons} not only exists among the antiferromagnetic members, it also persists in the most itinerant members of the U$M_2$Si$_2$ family.

\subsection{Singlets and quasi-doublets}
The symmetry of the $5f^2$ ground state is, according to our experiment, a singlet state so that the question arises how this is understood within the context of the antiferromagnetic ground states of UPd$_2$Si$_2$ and UNi$_2$Si$_2$ with very large ordered moments. After all, only the $\Gamma_5^{(1,2)}$ doublets carry a moment but none of the singlet states do (see Table\,\ref{Tab_moments}). The NIXS data, however, can also be described with two singlets states close in energy, i.e. with a quasi-doublet consisting of the $\Gamma_1^{(1)}$($\approx$90$^{\circ}$) and $\Gamma_2$ nearby in energy. Here the $\Gamma_1^{(1)}$ has a large $J_z$\,=\,$+$4 and $-$4 component and the $\Gamma_2$ is a pure $J_z$\,=\,$+$4 and $-$4 state and a quasi-doublet consisting of these two may carry an $induced$ moment. Actually, in the U$M_2$Si$_2$ structure, quasi-doublets consisting of $\Gamma_1^{(1,2)}$\,\&\,$\Gamma_2$ and of $\Gamma_3$\,\&\,$\Gamma_4$ are allowed by symmetry and the inter-site exchange of the $J_z$ components leads to the appearance of an ordered magnetic moment. Depending on the energy separation of the quasi-doublet and the admixture of the $J_z$ states in the molecular field ground state composed of $\Gamma_1^{(1,2)}$, any value between 0 and the maximum $J_z$ value may be reached. The range of magnetic moment values are listed in Table\,\ref{Tab_moments}. 
\begin{table}
    \centering
    \caption{Possible ordered magnetic moments $\mu_{\text{ord}}$ for the respective crystal-field states and of the quasi-doublets $\Gamma_1^{(1,2)}(\theta)$ \& $\Gamma_2$ and $\Gamma_3$ \& $\Gamma_4$ (third column).}
	  \label{Tab_moments}
\begin{tabular}{l|c|c}
\hline                          
    CEF states                 & $\mu_{\text{ord}}$ of singlet states & $\mu_{\text{ord}}$ of quasi-doublet \\  \hline \hline
$\Gamma_1^{(1,2)}(\theta)$     &      0     &  0-4$\mu_B$   \\ \cline{2-2} 
$\Gamma_2$                     &      0     &               \\ \hline
$\Gamma_3$                     &      0     &  0-2$\mu_B$   \\ \cline{2-2}  
$\Gamma_4$                     &      0     &               \\ \hline \hline
                               & $\mu_{\text{ord}}$ of doublet states &               \\ \hline
$\Gamma_5^{(1,2)}(\phi)$       & 0-3$\mu_B$ &               \\ \hline
  \end{tabular}
\end{table}

Our NIXS results reveal that the ground-state symmetry is that of the $\Gamma_1^{(1)}$ or the $\Gamma_2$ singlet, or the $\Gamma_1^{(1)}$\,\&\,$\Gamma_2$ quasi-doublet. Such a quasi-doublet can generate induced moments of up to 4\,$\mu_B$ (see Table\,1), thus naturally accommodating the large ordered moments that were observed in UPd$_2$Si$_2$\,\cite{Ptasiewicz1981,Shemirani1993} and UNi$_2$Si$_2$\,\cite{Lin1991}. Even the value of 3.37\,$\mu_B$\,\cite{Ptasiewicz1981} could be explained by such a quasi-doublet. We would like to point out that the idea of using a quasi-doublet to induce magnetic moments and long range AF magnetic order is not unrealistic. Such an induced magnetic moment scenario has been proposed to explain the magnetic moments in the moderate heavy-fermion compound UPd$_2$Al$_3$ with the dual nature of $f$ electrons explaining the heavy bands and dispersive magnetic singlet-singlet excitations mediating the superconducting pairing\,\cite{Sato2001,Thalmeier2002,Zwicknagl2003}. 

Another important aspect of having a quasi-doublet is that it allows for the degeneracy needed for a hidden order to occur in URu$_2$Si$_2$. Here, we argue that the orbital degrees of freedom rather than spin form the driving force for the phase transition. Furthermore, it should be mentioned that the Ising like anisotropy of the static susceptibility is compatible with a quasi-doublet consisting of these two states\,\cite{Nieuwenhuys1987}.

Interestingly, LDA+DMFT calculations found similar states, i.e., the local ground state and the first excited state of the U\,$f^2$ configuration in URu$_2$Si$_2$ are made up of $\Gamma_1^{(1)}$ and $\Gamma_2$ states (see Ref.\,\cite{Haule2009} and also the Supp. Mat. section of Ref.\,\cite{Kung2015}).  A complex Landau-Ginzburg theory based on these two states\,\cite{Haule2010} was developed for the HO and LAMFM phase of URu$_2$Si$_2$, and it accounts for the appearance of moments under applied pressure and other peculiarities of the HO phase. These calculations, however, do not include the full multiplet interactions, i.e., do not take into account the mixture of the orbital momenta $L$\,=3, 4, and 5.

\begin{figure}[bp]
    \centerline{\includegraphics[width=0.95\columnwidth]{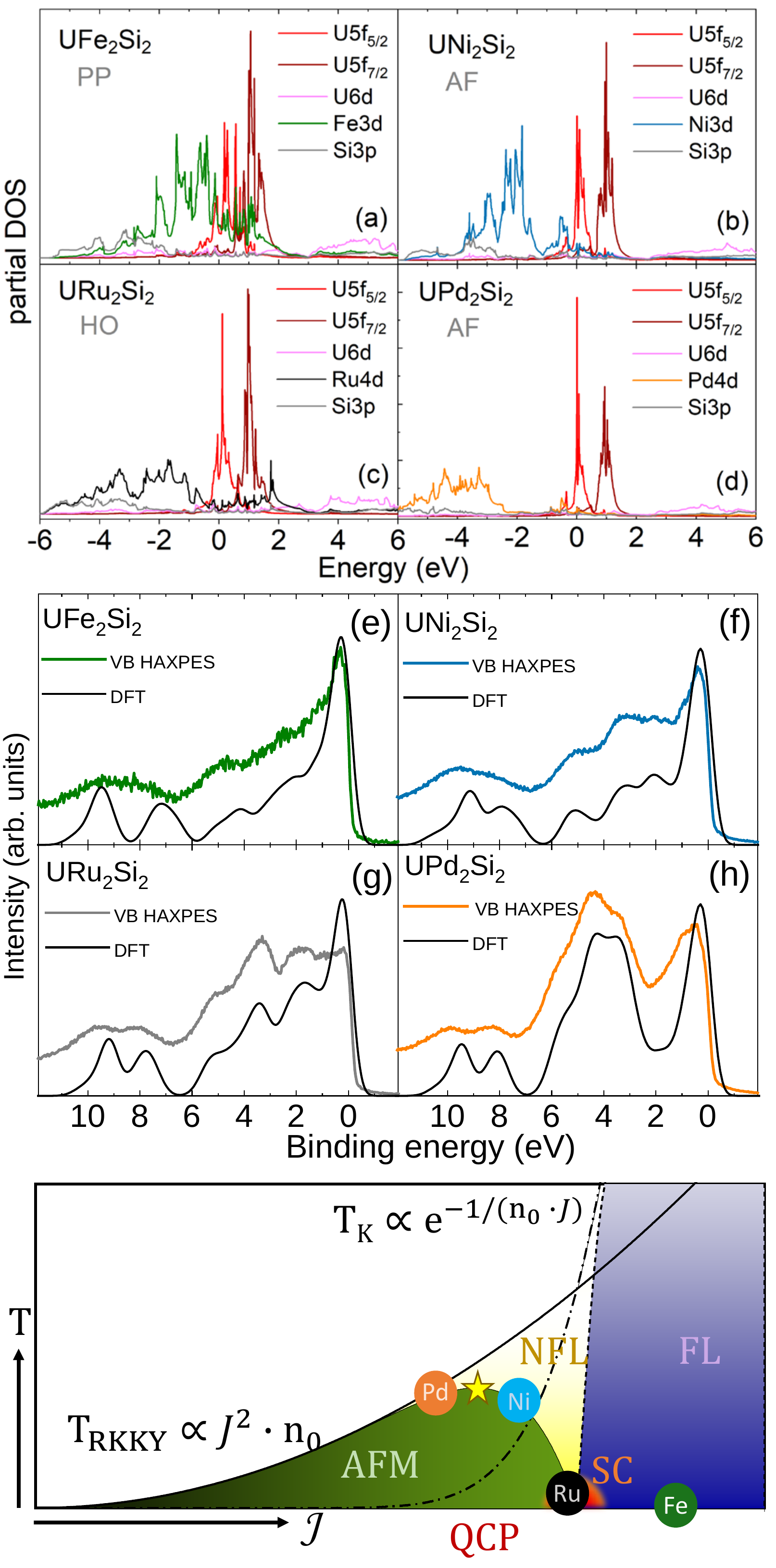}}
    \caption{(a)-(d) Partial DOS of U$M_2$Si$_2$ calculated with FPLO (see Appendix). The transition metal partial DOS of the $3d$ and $4d$ electrons are plotted in green (Fe), blue (Ni), black (Ru) and orange (Pd). (e)-(h) Experimental valence band (VB) HAXPES data of U$M_2$Si$_2$ compared to the DFT simulated spectra, which have been obtained from the calculated uranium, transition metal, and silicon partial DOS'ses weighted for the respective shell-specific photo-ionization cross-sections. The incident energy was h$\nu$\,=\,5945\,eV. (i) Doniach-like phase diagram of U 5$f^2$ within the quasi-doublet scenario: temperature $T$ versus exchange interaction $\cal{J}$ diagram showing the antiferromagnetic regime (AF, green), the intermediate valent Fermi liquid (FL, purple), the non-Fermi liquid (NFL, yellow), the superconducting dome (SC, orange) close to the quantum critical point (QCP). The dots represent the location of the respective members of the U$M_2$Si$_2$. $T_K$ refers to the Kondo-like temperature and $T_{\text{RKKY}}$ to the Ruderman-Kittel-Kasuya-Yosida temperature scale. The yellow star that marks the maximum Neel temperature corresponds to ($\cal{J}$$_{max}$,$T_{max}$)\,$\approx$\,(0.27,0.01) and the critical value is $\cal{J}$$_c$(QCP)\,$\approx$\,0.36 in units of the conduction band width W\,$\sim$\,1/n$_0$ where n$_0$ is the DOS\,\cite{Schafer2019}. }
    \label{DOS}
\end{figure}

\subsection{$f$-$d$ hybridization strength}
We need to look at the hybridization process between the $5f$ and the conduction band electrons in order to explain the increase of the U$^{3+}$\,$5f^3$ spectral weight in the sequence $M$\,=\,Pd(AF)\,$\rightarrow$\,Ni(AF)\,$\rightarrow$\,Ru(HO)\,$\rightarrow$\,Fe(PP), 

Figure\,\ref{DOS}\,(a)-(d) show the result of density functional theory (DFT) calculations in the non-magnetic phase using FPLO\,\cite{Koepernik1999} (see Appendix). The partial density of states (DOS) of U 5$f$ for $J$\,=\,5/2 and 7/2, the U 6$d$, the transition metal 3$d$ or 4$d$, respectively, and the Si 3$p$ partial DOS are displayed. We observe first of all that there are transition metal $d$ states (colored green (Fe), black (Ru), blue (Ni), and orange (Pd)) present in the energy region where the U $5f_{5/2}$ (colored red) is located, i.e. around the Fermi level. The amount is appreciable for the Fe compound, and gets smaller for the Ru and Ni, and is tiny for the Pd. In addition, a closer look reveals that the width of the U $5f_{5/2}$ band is the largest for the Fe compound and the smallest for the Pd, with the Ru and Pd in between. All together, this indicates that the mixing or hybridization between the transition metal $d$ and the U $5f$ states is the strongest for Fe, and decreases for the Ru and Ni, and is the weakest for the Pd. This trend is fully consistent with the U $4f$ HAXPES result in that the U 5$f^3$ contribution decreases successively from the Pauli paramagnet UFe$_2$Si$_2$ via the hidden order compound URu$_2$Si$_2$ to the two antiferromagnets UNi$_2$Si$_2$  and UPd$_2$Si$_2$.

For the interpretation of the calculated partial DOS, one can look whether the hybridization trend of Fe-Ru-Ni-Pd is reflected by the U-U distances $a$ or the U-transition metal distances $d_{\text{U-TM}}$ across this set of compounds. If so, the $f$-$d$ hopping integral may play an important role. However, the trend of $a$, from small to large, goes as Fe, Ni, Ru and Pd (see, e.g., Fig\,\ref{table} in Appendix). The trend for $d_{\text{U-TM}}$ is, from small to large, Ni, Fe, Ru and Pd. In other words, already the opposite order of the Ni and Ru compounds in both, $a$ and $d_{\text{U-TM}}$ does not favor an interpretation of hopping integral driven systematics. What is clearly forming a trend in the calculations is the energy position of the $d$ states relative to the U $5f$. In going from $M$\,=\,Fe to Ru to Ni, and to Pd, we observe that the $d$ states are moving to more negative energies and thus farther away from the U $5f$ levels. This strongly suggests that, in terms of a many body model such as an Anderson impurity or lattice model, it is the $5f$ level position $\epsilon_f$ that provides the crucial parameter when comparing this U$M_2$Si$_2$ series. 

Valence band HAXPES has been used to check whether these DFT predictions are valid. Figure\,\ref{DOS}\,(e)-(h) shows the spectra of the four compounds. In order to compare the DFT results to the experiment, we calculate the valence band spectra by multiplying each of the partial DOS by their respective shell-specific photoionization cross-section at 5945\,eV photon energy as derived from from Refs.\,\cite{TRZHASKOVSKAYA2006245} and by the Fermi function to include only the contributions from the occupied states, followed by a broadening to account for the experimental resolution and intrinsic broadening, and their summation. This was done for all the partial DOS included in the calculation (not only the ones shown in Figure\,\ref{DOS}\,(a)-(d)). The results are displayed in Figure\,\ref{DOS}\,(e)-(h) (black lines). The comparison with valence band HAXPES data confirms the validity of the DFT predictions; the general agreement between experiment and theory is very good. There are deviations in the U $5f$ regions where the calculated intensities are higher than in the experiment since correlation effects were neglected in the DFT, but the line shape and positions of the silicon, transition metal, and uranium non-$f$ bands are well reproduced. This validates the hybridization picture offered by the DFT calculations, and in turn, provides a consistent reasoning for the increase of the $5f^3$ spectral weight in the sequence $M$\,=\,Pd(AF)\,$\rightarrow$\,Ni(AF)\,$\rightarrow$\,Ru(HO)\,$\rightarrow$\,Fe(PP).

\subsection{Dual nature of the $5f$ electrons}
The above findings are very much compatible with the dual-nature idea of $f$ electrons in uranium heavy fermion compounds\,\cite{Zwicknagl2002,Zwicknagl2003a,Thalmeier2002,Zwicknagl2003}. On one hand, we have observed in NIXS the local atomic multiplet structure of the U\,$5f^2$ configuration in the U$M_2$Si$_2$ system. On the other hand, we have noticed from HAXPES the intermediate valent character of U in U$M_2$Si$_2$, and that the U\,$5f^3$ weight increases from Pd to Ni to Ru to Fe. Thus, with increasing $f$-$d$ hybridization the magnetic moments get suppressed until eventually an intermediate valent Fermi liquid state with enhanced Pauli paramagnetism is reached, thereby showing the impact of the itinerant part. Two of the $5f$ electrons remain localized and form atomic multiplet states whereas a third electron is effectively delocalized with an accordingly renormalized mass.

An important finding is that the four compounds share the same multiplet states for the 5$f^2$ configuration, namely the $\Gamma_1^{(1)}$($\approx$90$^{\circ}$)/$\Gamma_2$ quasi-doublet. This allows us to draw a Doniach-like phase diagram in which the temperature $T$ is plotted versus an effective exchange interaction $\cal{J}$, a quantity that is determined by the $f$-$d$ hopping integral $V$ and the $f$ energy $\epsilon_f$\,\cite{Khomskii2010} and that can be associated with the degree of delocalization of the third electron. For small $\cal{J}$ magnetic order prevails, whereas for large $\cal{J}$ a Kondo-like screened (intermediate valent) state forms that is well described in terms of a Fermi liquid (FL) with enhanced Pauli paramagnetism. In the transition region a quantum critical point (QCP) and non-Fermi liquid (NFL) scaling occurs that is often hidden by a superconducting ($SC$) dome\,\cite{Hilbert2007,Wirth2016}. In Fig.\,\ref{DOS}\,(i) the Pd member of the family is placed the most to the left because it has the largest ordered moment, followed by the Ni compound that also resides in the AF regime. URu$_2$Si$_2$, however, is placed very close to or at the QCP since it is the only compound of the family that exhibits superconductivity and hidden order. UFe$_2$Si$_2$, finally, is located on the Kondo-like screened side (PP) to the right of the QCP where the physical properties follow FL scaling. 

The application of pressure is known to push URu$_2$Si$_2$ into the AF regime\,\cite{Amitsuka2007,Niklowitz2010}, i.e., pressure reduces the itinerant part, and in our picture, reduces $\cal{J}$. This may seem counterintuitive since pressure will decrease distances and hence increase $V$ so that $\cal{J}$ becomes larger since it is proportional to $V^2$/$\epsilon_f$\,\cite{Khomskii2010}. However, we know pressure will stabilize the $f^2$ configuration with its smaller ionic radius at the expense of the $f^3$. This is reflected in this description by the increase of $\epsilon_f$ whereby $\epsilon_f$ is positive to denote that the $f^2$ configuration is lower in energy than the $f^3$. Hence, $\cal{J}$ is decreased with pressure because $\epsilon_f$ increases more strongly than $V^2$. With this in mind, we speculate applying pressure to UFe$_2$Si$_2$ will also reduce $\cal{J}$. And indeed, DFT calculations for a compressed UFe$_2$Si$_2$ lattice\,\cite{Kuwahara2003} find that the $f^2$ configuration is increasingly populated as pressure rises (see Fig.\,\ref{population} in Appendix), thus moving the Fe compound closer to the superconducting-hidden order regime.  

We would like to note that already in 1993, Endstra \textsl{et al}.\,\cite{Endstra1993b} sorted the members of the U$M_2$Si$_2$ family into a Doniach-like phase diagram. The same sequence was suggested, but this was merely based on semiquantitative band structure calculations of the hybridization strength. Which local atomic like states are active were not known, and moreover, the issue whether the U$M_2$Si$_2$ members have the same multiplet states in common was not even considered. Our experimental findings justify the use of a Doniach-like phase diagram since a common quasi-doublet scenario can be established for the local states together with the observation of strongly varying $5f$ count across the family. Of utmost importance is the fact that the particular quasi-doublet scenario made of $J$\,=\,4 states allows for the large span of properties across the U$M_2$Si$_2$ family, namely to cover antiferromagnetism with very large ordered moments, hidden order and superconductivity, as well as Pauli paramagnetism.

\section{Conclusion}
The dual nature of the 5$f$ electrons in four isostructural compounds with very different ground state properties, namely UPd$_2$Si$_2$ (AF), UNi$_2$Si$_2$ (AF), URu$_2$Si$_2$ (HO), and UFe$_2$Si$_2$ (PP) has been shown. The NIXS data of the U $O_{4,5}$ edge reveal multiplets of the localized U\,5$f^2$ configuration in all four compounds, irrespective of the degree of itineracy, and the directional dependence of NIXS unveils that the different collectively ordered (or non-ordered) ground states form out of the same symmetry. The symmetry is determined by the singlet states $\Gamma_1^{(1)}$($\approx$90$^{\circ}$) or $\Gamma_2$ of the U\,5$f^2$ Hund's rule ground state, so that only an induced-type of order with a quasi-doublet consisting of these two singlet states can explain the large ordered moment of the antiferromagnetic members of the family. The comparison of the 4$f$ core-level HAXPES data is meaningful because the four compounds have the same local ground-state symmetry. It reveals the change of the itinerant character within the family. The $relative$ 5$f$-shell filling increases successively when going from $M$\,=\,Pd(AF)\,$\rightarrow$\,Ni(AF)\,$\rightarrow$\,Ru($HO$)\,$\rightarrow$\,Fe(PP) so that a comprehensive picture is proposed, namely the sorting of the U$M_2$Si$_2$ compounds into a Doniach-like phase diagram.

\section{Appendix}

\subsection{Sample Preparation}
The URu$_2$Si$_2$ single crystals used for HAXPES were grown by the Czochralski method in a tetra-arc furnace in San Diego from high purity starting elements (depleted uranium\,--3N, Ru\,--3N, Si\,--6N). Single crystalline URu$_2$Si$_2$ used for the NIXS experiment was grown with the traveling zone method in the two-mirror furnace in Amsterdam under high-purity (6N) argon atmosphere. Single crystals of U$M_2$Si$_2$ with $M$ = Fe, Ni, and Pd were grown in Wroclaw by Czochralski pulling technique in ultra-pure Ar atmosphere using a tetra-arc furnace. The starting components were high-purity elements (natural uranium\,--\,3N, Fe\,--\,3N, Ni\,--\,4N, Pd\,--\,4N, and Si\,-- 6N). All single crystals were checked x-ray Laue diffraction for thies single-crystalline nature.

Polycrystalline UPd$_3$ sample of 1\,g was synthesized in Dresden by arc melting stoichiometric amounts of uranium metal (natural, foil, Goodfellow, 99.98 wt.\%) with palladium metal (shot, Chempur, 99.99 wt.\%) under a protective atmosphere of argon gas. The melted button was then placed into an alumina crucible and sealed into a tantalum tube. The sample was heated to 1400$^{\circ}$C within 6\,hours, annealed for additional 6\,hours and subsequently furnace cooled to room temperature. The single phase nature of the sample was deduced from the analysis of powder XRD data.

Single crystals of UCd$_{11}$ were grown from Cd flux in Los Alamos.  Uranium and cadmium pieces in the molar ratio U:Cd\,=\,1:133 were placed in an alumina crucible and sealed under vacuum in a silica ampoule.  The ampoule was heated to 600$^{\circ}$\,C, held at that temperature for 20\,hours, then slowly cooled at 2$^{\circ}$\,C/hr. to 400$^{\circ}$\,C, whereupon the excess Cd flux was removed via a centrifuge.

\subsection{Experiment}
The NIXS measurements were performed at the High-Resolution Dynamics Beamline P01 of the PETRA-III synchrotron in Hamburg, Germany. The end station has a vertical geometry with twelve Si(660) 1\,m radius spherically bent crystal analyzers that are arranged 3$\times$4 matrix and positioned at scattering angles of 2\,$\theta$\,$\approx$\,150$^\circ$, 155$^\circ$, and 160$^\circ$. The final energy was fixed at 9690\,eV, the incident energy was selected with a Si(311) double monochromator, and the overall energy resolution was $\approx$\,0.7\,eV. The scattered beam was detected by a position sensitive custom-made Lambda detector based on a Medipix3 chip. A sketch of the scattering geometry can be found in Ref.\,\cite{Sundermann2017}. The averaged momentum transfer was $|\vec{q}|$\,=\,(9.6\,$\pm$\,0.1)\,\AA$^{-1}$ at the U $O_{4,5}$ edge. The crystals were mounted in a Dynaflow He flow cryostat with Al-Kapton windows.

The HAXPES experiments were carried out at the beamlines P09 and P22 of the PETRA-III synchrotron in Hamburg, Germany\,\cite{Strempfer2013,Schlueter2019}. The incident photon energy was set at 5945\,eV. The valence band spectrum of a gold sample was measured in order to determine the Fermi level $E_F$ and the overall instrumental resolution of 300\,meV. The excited photoelectrons were collected using a SPECS225HV electron energy analyzer in the horizontal plane at 90$^{\circ}$. The sample emission angle was 45$^{\circ}$. Clean sample surfaces were obtained by cleaving the samples \textsl{in situ} in the cleaving chamber prior to inserting them into the main chamber where the pressure was $\sim$10$^{-10}$\,mbar. The measurements were performed at a temperature of 20\,K. 

\subsection{Simulation}
The simulations include the spin-orbit as well as Coulomb interactions with atomic values from the Cowan code. The Slater integrals 5$f$-5$f$ and 5$d$-5$f$ were reduced in order to account for configuration interaction\,\cite{Tanaka1994} and covalency effects\,\cite{Agrestini2017} that are not included in the Hartree-Fock scheme. A reduction of 50\% reproduces the energy distribution of the multiplet excitations of the U $O_{4,5}$-edges of the U$M_2$Si$_2$. As in Ref\,\cite{Sundermann2016}, the ratio of multipoles was slightly adjusted by using a $|\vec{q}|$ value that is slightly larger than the experimental one. This is necessary because the radial wave functions are based on atomic values. The $J$\,=\,4 multiplet forms the ground state for all finite values of spin-orbit coupling and Coulomb interaction. The relative contributions of the orbital angular momenta $L$\,=\,3, 4, and 5 are 1\%, 14\%, and 85\% for the present ratio of spin-orbit coupling and Coulomb interaction. A Gaussian broadening of 0.7\,eV accounts for the instrumental resolution and a Lorentzian broadening of 1.3\,eV for life-time effects. In addition some asymmetry due to the metallicity of the samples has been described by using a Mahan-type line shape with an asymmetry factor of 0.18 and an energy continuum of 1000\,eV. 

\subsection{DFT calculation}
Density functional theory based calculations were performed using the full-potential non-orthogonal local orbital code (FPLO v.18.00.52) employing the local density approximation (LDA) and including spin-orbit coupling (fully relativistic calculation). A grid of 15x15x15 k-points and 5000 energy points (about 1 point every 8\,meV) have been used for the calculation of the band structure and density of states (DOS).  

\begin{figure}[]
 	 \centerline{\includegraphics[width=0.9\columnwidth]{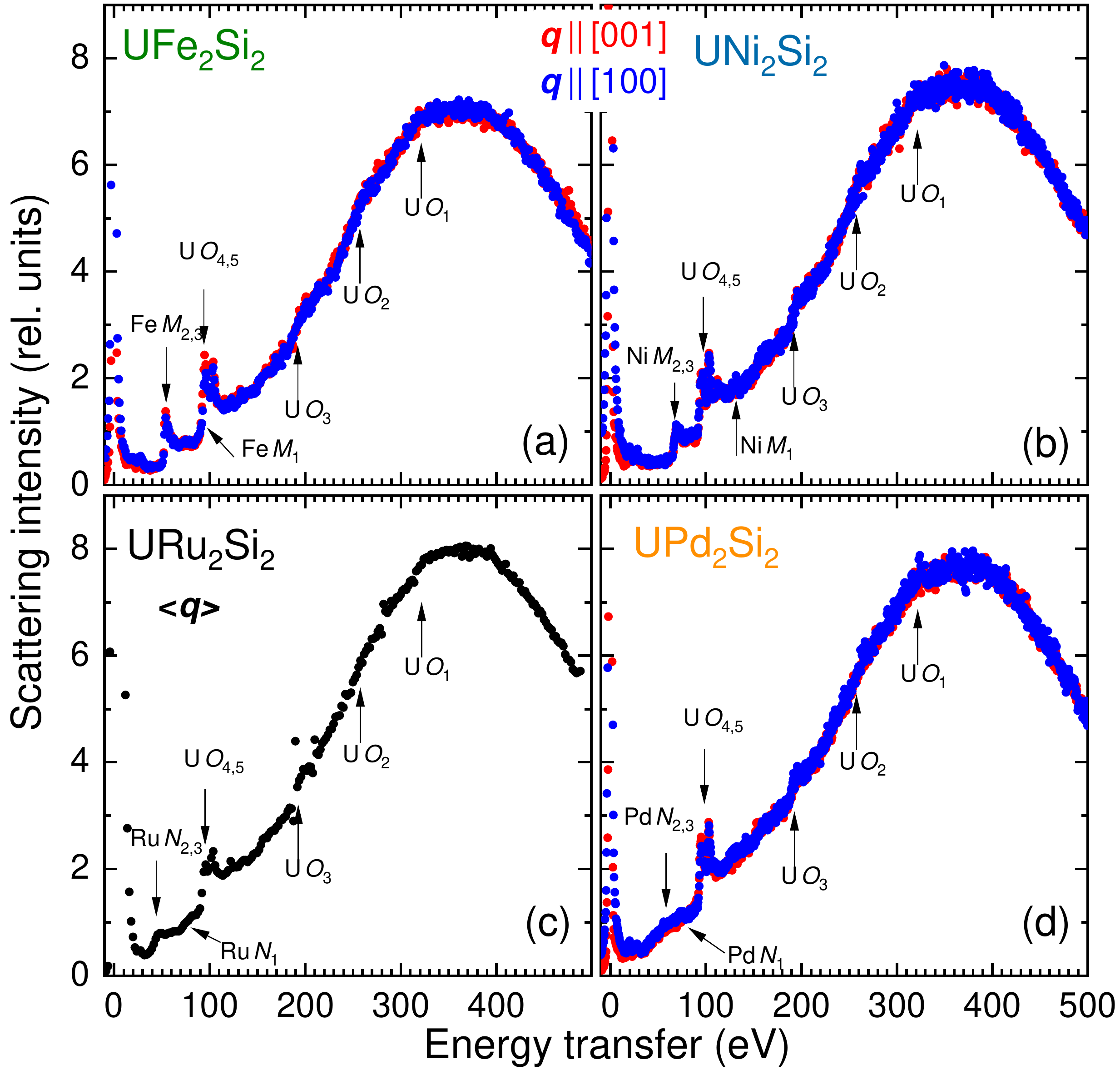}}
    \caption{ Experimental wide energy scans at $T$\,$<$\,20\,K covering the peak of the Compton signal of (a) UFe$_2$Si$_2$, (b) UNi$_2$Si$_2$, (c) URu$_2$Si$_2$ and of (d) UPd$_2$Si$_2$; red and blue denote the two different directions of the momentum transfer $\vec{q}$. The data of different momentum directions are scaled to the Compton peak. The URu$_2$Si$_2$ data in panel (c) were acquired in a previous experiment on a different beamline with a larger step size\,\cite{Sundermann2016,Sundermann_PhD} and were averaged over several directions of $\vec{q}$.}
    \label{Compton}
\end{figure}

\subsection{Compton}
 Figure\,\ref{Compton}(a)-(d) show the experimental NIXS spectra of UFe$_2$Si$_2$ (a), UNi$_2$Si$_2$ (b), URu$_2$Si$_ 2$(c) and UPd$_2$Si$_2$ (d) over a wide energy range with a coarse energy step size of 0.5\,eV (even larger for URu$_2$Si$_2$) for different directions of the momentum transfer $\vec{q}$, $\vec{q}$$\|$[100] (blue) and $\vec{q}$$\|$[001] (red). The URu$_2$Si$_2$ data were obtained in a previous experiment (see Ref.\,\cite{Sundermann2016,Sundermann_PhD}) and here only $<$$\vec{q}$$>$ averaged signal is shown. The dominant signal arises from Compton scattering and the core-level excitations appear as spikes on top.
 
\subsection{Isotropic NIXS spectra of U\,$f^2$ and U\,$f^3$}
\begin{figure}[]
	  \centerline{\includegraphics[width=0.9\columnwidth]{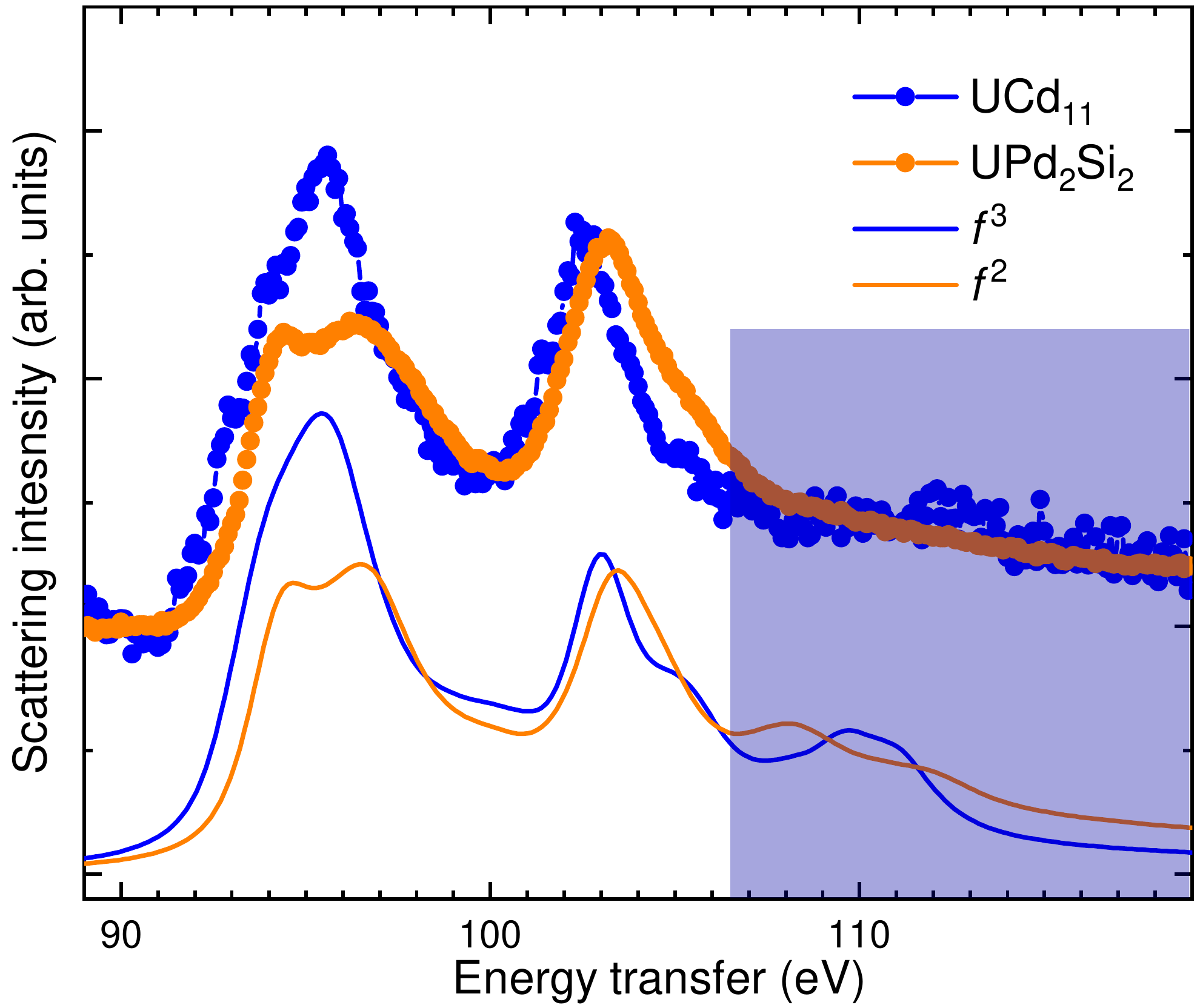}}
   \caption{Isotropic NIXS data of UPd$_2$Si$_2$ (orange dots) and of UCd$_{11}$ (blue dots) together with the simulations for a pure U\,$f^2$ (orange lines) and U\,$f^3$ (blue lines) configurations. The shaded area indicates the energy regime where excitations to the continuum states are becoming more intense.  }
    \label{isotropic}
\end{figure}

Figure\,\ref{isotropic} shows the experimental isotropic O4,5 NIXS spectra of UCd$_{11}$ (blue dots) and of UPd$_2$Si$_2$ (orange dots), the former very much on the $f^3$ side and the latter on the $f^2$. The line shapes are very different for the different U configurations. In addition, we observe that the UCd$_{11}$ spectrum is shifted towards lower energies by about 0.5-1.0 eV which is typically observed in x-ray absorption studies when comparing configurations with valence states that differ by one; with the lower valence state lower in energy, see e.g. S. Agrestini \textit{et al.} \,\cite{Agrestini2018} and references there in. 

Figure\,\ref{isotropic} also displays the theoretical simulations for the isotropic O4,5 NIXS spectra from a pure local U\,$f^3$ (blue lines) and  
local U\,$f^2$ (orange lines) configurations. We can observe that the simulations reproduce very well the experimental spectra, and therefore we can safely infer that the respective configuration assignments of the two compounds are correct. The shaded area denotes the energy range where excitations to the continuum states are becoming more intense. The interference of the atomic-like transitions with these continuum states causes considerable broadening of the spectral features, see e.g. Caciuffo \textit{et al.} \cite {Caciuffo2010}, which has not been included in the simulations.   

\subsection{Properties}
\begin{figure}[htb]
	  \centerline{\includegraphics[width=0.8\columnwidth]{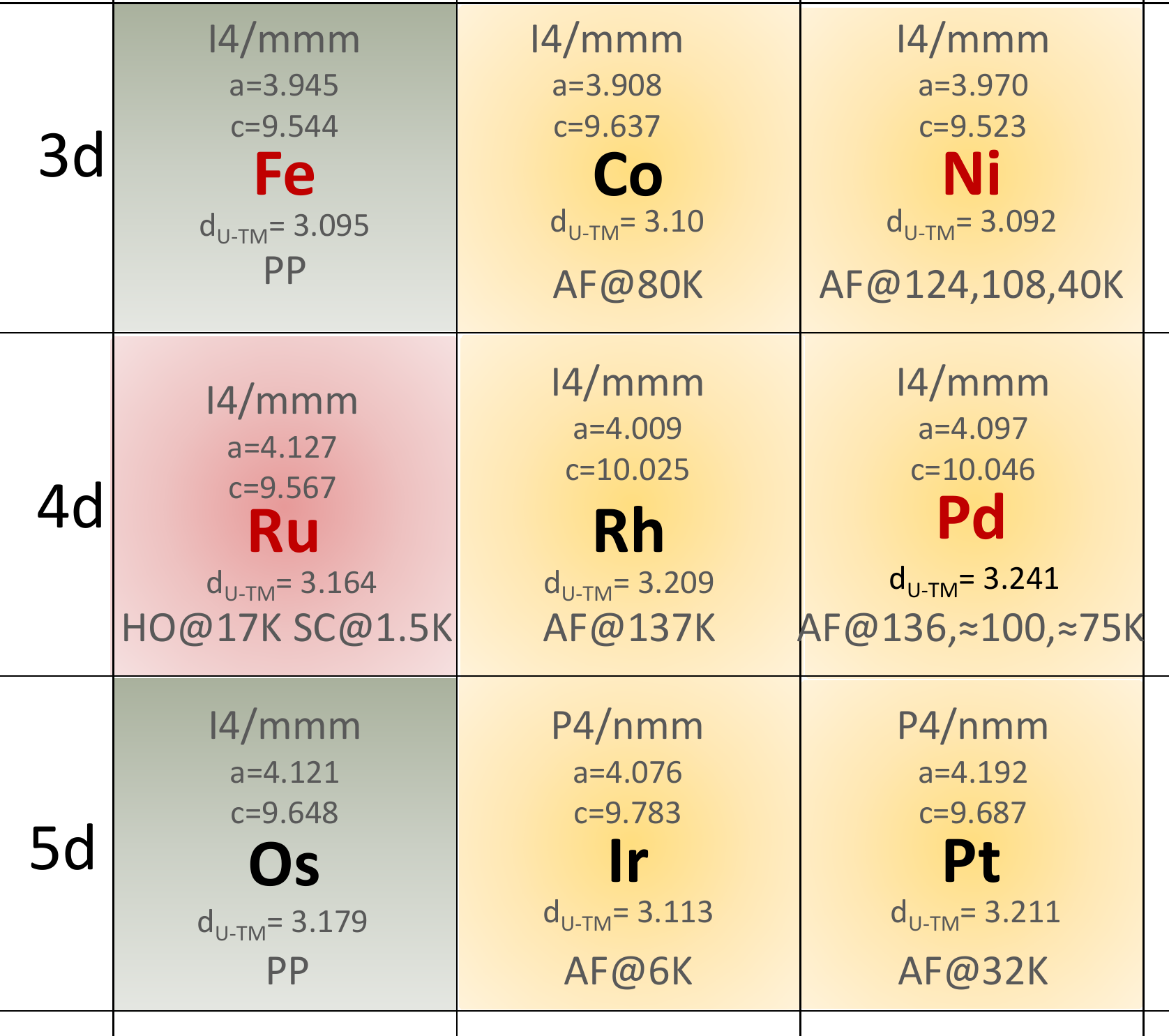}}
   \caption{Section of the periodic table for the U$M_2$Si$_2$ compounds providing the crystallographic structure, the lattice constants, the U--transition metal distances, the ground state properties AF for antiferromagnetic, PP for Pauli paramagnetic, HO for hidden order, SC for superconducting, as well as the temperatures at which the respective transitions take place\,\cite{Ptasiewicz1981,Buschow1986,Palstra1986,Endstra1993b,Shemirani1993,Svoboda2004,Szytuka2007,Plackowski2011}.}
    \label{table}
\end{figure}

Figure\,\ref{table} summarizes the lattice constants, U transition metal (TM) distances, structure, and ground state properties of the U$M_2$Si$_2$ compounds surrounding URu$_2$Si$_2$. 

\subsection{DFT of UFe$_2$SI$_2$ under pressure}
\begin{figure}[htb]
	  \centerline{\includegraphics[width=0.9\columnwidth]{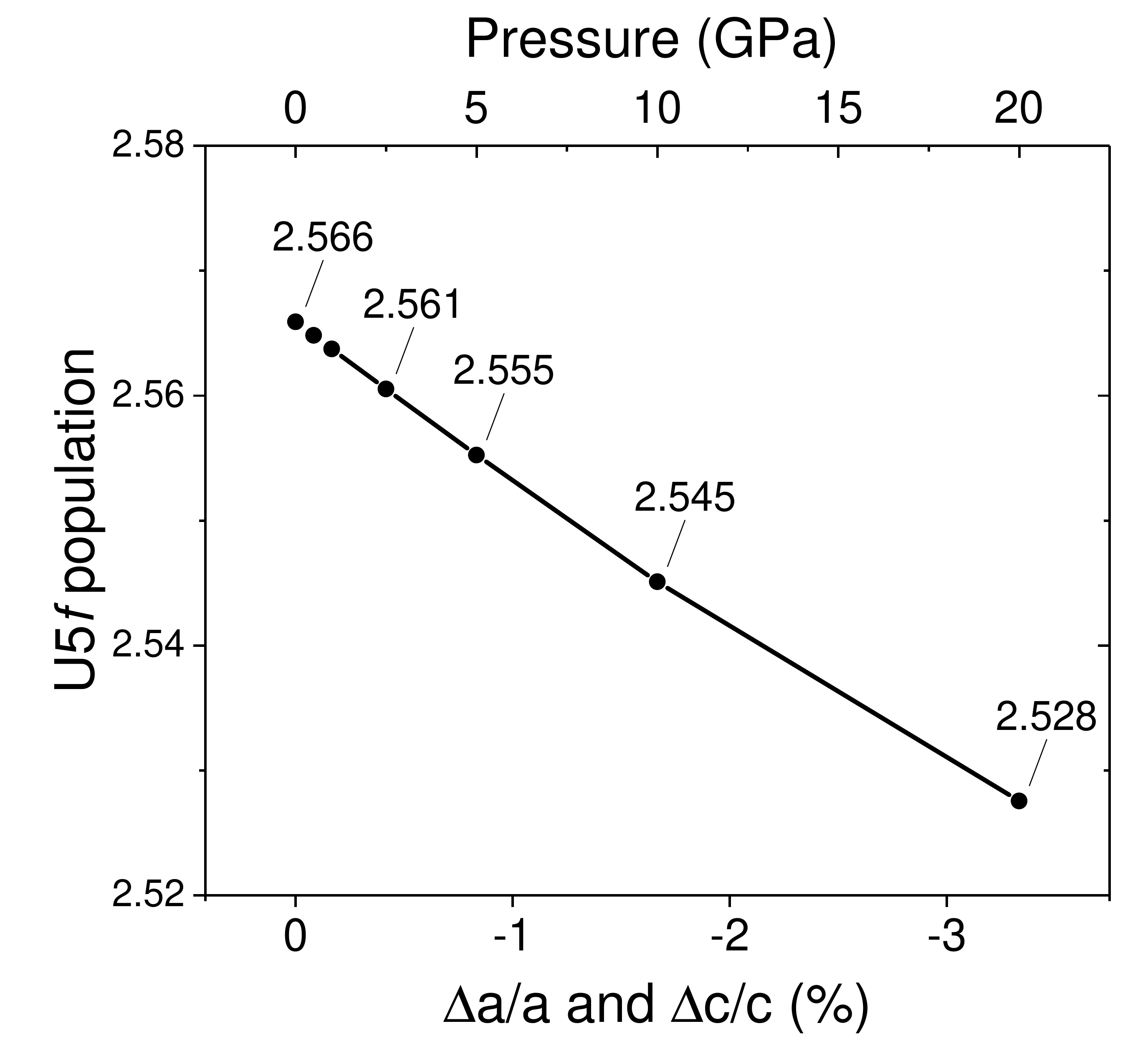}}
    \caption{U\,5$f$ population in UFe$_2$Si$_2$ as function of $\Delta$a/a and $\Delta$c/c in \% resulting from DFT calculation using FPLO. The conversion to pressure is based on the compressibility data of URu$_2$Si$_2$\,\cite{Kuwahara2003}. }
    \label{population}
\end{figure}

Figure\,\ref{population} shows the result of DFT calculations for UFe$_2$Si$_2$ assuming reduced lattice constants to mimic the impact of applied pressure. The conversion of reduced lattice constants to pressure is done on the basis of compressibility data of URu$_2$Si$_2$\,\cite{Kuwahara2003} since data for UFe$_2$Si$_2$ are not available.  The calculations clearly show that the \textit{relative} 5$f$ population decreases with increasing pressure, i.e. the 5$f^3$ contributes less to the ground state as pressure increases. (Note the \textit{absolute} 5$f$ occupation numbers should be taken with care  since they depend on the calculational basis set and method of counting.)

\section{Acknowldegement}
This research were carried out at PETRA\,III/DESY, a member of the Helmholtz Association HGF. A.  Amorese, A. Severing, and M. Sundermann gratefully acknowledge the financial support of the Deutsche Forschungsgemeinschaft under project SE\,1441-5-1. M. Szlawska was supported by the National Science Centre of Poland, Grant No. 2018/31/D/ST3/03295. Research at UC San Diego was supported by the US Department of Energy, Office of Basic Energy Sciences, Division of Materials Sciences and Engineering, under Grant No. DEFG02-04-ER46105 (single crystal growth) and US National Science Foundation under Grant No. DMR-1810310 (materials characterization). Work at Los Alamos National Laboratory was performed under the auspices of the US DOE, OBES, DMSE. All authors thank H. Borrman from MPI-CPfS for his support in handling the U based samples and J. Grin for support and interest.

%

\end{document}